\documentclass[aps,longbibliography,
  twocolumn,amsmath,
  amssymb,nofootinbib, superscriptaddress, preprintnumbers]{revtex4-1}

\usepackage{graphics}      
\usepackage{graphicx}      
\usepackage{longtable}     
\usepackage{url}           
\usepackage{bm}            
\usepackage{xcolor}
\usepackage{mathrsfs}
\usepackage[colorlinks,allcolors=blue]{hyperref}
\usepackage{blindtext}
\usepackage{hyperref}
\usepackage{float}
\usepackage{bbold}
\usepackage{lipsum}
\usepackage[normalem]{ulem}
\usepackage{orcidlink}
\usepackage{academicons}
\usepackage{xcolor}
\usepackage{subfigure}
\usepackage{booktabs}
\usepackage{hyperref}

\newcommand{\orcid}[1]{\href{https://orcid.org/#1}{\textcolor[HTML]{A6CE39}{\aiOrcid}}}

\begin{document}
\preprint{MPP-2023-271}

\newcommand{\Sajad}[1]{\textcolor{red}{\textbf{Sajad:} #1}}
\newcommand{\Akira}[1]{\textcolor{blue}{\textbf{Akira:} #1}}

\title{Machine Learning-Based Detection of Non-Axisymmetric Fast Neutrino Flavor Instabilities in Core-Collapse Supernovae}

\newcommand*{\MPP}{\textit{\small{Max-Planck-Institut f\"ur Physik (Werner-Heisenberg-Institut), F\"ohringer Ring 6, 80805 M\"unchen, Germany}}}

\author{Sajad Abbar \orcidlink{0000-0001-8276-997X}   } 
\affiliation{\MPP}
\author{Akira Harada} 
\affiliation{RIKEN Interdisciplinary Theoretical and Mathematical Sciences Program (iTHEMS), 2-1 Hirosawa, Wako, Saitama 351-0198, Japan}
\author{Hiroki Nagakura \orcidlink{0000-0002-7205-6367}} 
\affiliation{Division of Science, National Astronomical Observatory of Japan, 2-21-1 Osawa, Mitaka, Tokyo 181-8588, Japan}

\begin{abstract}
In dense neutrino environments like core-collapse supernovae (CCSNe) and neutron star mergers (NSMs), neutrinos can undergo fast flavor conversions (FFC) when their angular distribution of neutrino electron lepton number ($\nu$ELN) crosses zero along some  directions. While previous studies have demonstrated the detection of axisymmetric $\nu$ELN crossings in these extreme environments, non-axisymmetric crossings have remained elusive, mostly due to the absence of models for their  angular distributions. In this study, we present a pioneering analysis of the detection of non-axisymmetric $\nu$ELN crossings using machine learning (ML) techniques. Our ML models are trained on  data from two CCSN simulations, one with rotation and one without, where non-axisymmetric features in neutrino angular distributions play a crucial role. We demonstrate that our ML models achieve  detection accuracies exceeding 90\%.
This is an important improvement, especially considering that a significant portion of $\nu$ELN crossings in these models eluded detection by earlier methods.
 \end{abstract}

\maketitle

\section{Introduction}

Core-collapse supernovae (CCSNe) and neutron star mergers (NSMs) are among the most extreme astrophysical settings in the universe. 
These events, marking the dramatic finale of massive stars and the collision of densely packed remnants, reveal some of the universe's most energetic and enigmatic events~\cite{Burrows:2020qrp, Janka:2012wk, Foucart:2022bth, Kyutoku:2021icp}.

At the heart of these celestial dramas lies  the emission of elusive neutrinos, which are generated in huge numbers during both CCSNe and NSMs.
In the extreme and densely packed conditions within these events, neutrinos undergo  an exciting journey marked by an intriguing phenomenon referred to as collective neutrino oscillations~\cite{pantaleone:1992eq, sigl1993general, Pastor:2002we,duan:2006an, duan:2006jv, duan:2010bg, Mirizzi:2015eza}~(for a recent review see Ref.~\cite{Volpe:2023met}). This fascinating behavior emerges from their coherent forward scatterings  with the dense  neutrino background. This nonlinear and collective dance gives rise to a rich tapestry of flavor transformations.

One of the most interesting  advancements in the field has been the discovery and  the exploration  of the fascinating phenomenon of
 \emph{fast} flavor conversions (FFCs), which occurs on extraordinarily short  scales (see, e.g., Refs.~\cite{Sawyer:2005jk, Sawyer:2015dsa,
Chakraborty:2016lct, Izaguirre:2016gsx,Capozzi:2017gqd, Abbar:2017pkh, Abbar:2018beu,Capozzi:2018clo, Martin:2019gxb, Abbar:2018shq, Abbar:2019zoq, Capozzi:2019lso, Johns:2019izj, Martin:2021xyl, Tamborra:2020cul,  Sigl:2021tmj, Kato:2021cjf,  Morinaga:2021vmc, Nagakura:2021hyb,  Sasaki:2021zld, Padilla-Gay:2021haz, Abbar:2020qpi, Capozzi:2020syn, DelfanAzari:2019epo, Harada:2021ata,  Abbar:2021lmm, Just:2022flt, 
Padilla-Gay:2022wck, Capozzi:2022dtr, Zaizen:2022cik,  Kato:2022vsu, Zaizen:2022cik,  Bhattacharyya:2020jpj, Wu:2021uvt, Richers:2021nbx, Richers:2021xtf, Dasgupta:2021gfs, Nagakura:2022kic, Ehring:2023lcd, Ehring:2023abs, Xiong:2023vcm, Zaizen:2023ihz, Xiong:2023upa, Fiorillo:2023hlk, Nagakura:2023wbf, Martin:2023gbo, Fiorillo:2023mze,
Grohs:2023pgq, Abbar:2020fcl, Johns:2021taz, Grohs:2023pgq, Cornelius:2023eop, Froustey:2023skf, Fiorillo:2024fnl, Abbar:2023ltx, Richers:2022bkd, George:2022lwg, Wu:2021uvt}). 
 FFCs manifest on scales  characterized by $\sim G_{\rm{F}}^{-1} n_{\nu}^{-1}$, potentially spanning just a few centimeters within the  SN core. This is in contrast to the conventional slow modes, governed by the neutrino vacuum frequency and expected to extend over kilometer-scale distances within the SN environment.
 Here, $G_{\rm{F}}$ represents the Fermi coupling constant, and $n_{\nu}$ denotes the neutrino number density.

FFCs occur \emph{iff} the angular distribution of the neutrino electron lepton number, $\nu$ELN, 
\begin{equation}
  G(\mathbf{v}) =
  \sqrt2 G_{\mathrm{F}}
  \int_0^\infty  \frac{E_\nu^2 \mathrm{d} E_\nu}{(2\pi)^3}
        [ f_{\nu_e}(\mathbf{p}) -  f_{\bar\nu_e}(\mathbf{p})],
 \label{Eq:G}
\end{equation}
crosses zero at some $\mathbf{v} = \mathbf{v}(\mu,\phi_\nu)$, with $\mu =\cos\theta_\nu$~\cite{Morinaga:2021vmc}. 
Here, $E_\nu$, $\theta_\nu$, and $\phi_\nu$ are the neutrino energy,  
the zenith, and azimuthal angles of the neutrino velocity, respectively, 
and  $f_{\nu}$'s are the neutrino 
occupation numbers. Note that here and throughout this work we assume that 
$\nu_x$ and $\bar\nu_x$ (heavy-lepton neutrinos and antineutrinos) have 
{\bf the same}
angular distributions.

 Investigating $\nu$ELN crossings requires access to comprehensive angular distributions of neutrinos. Yet, acquiring this detailed angular data presents a considerable challenge in modern CCSN  and NSM simulations due to the extensive computational resources it demands.
In light of this challenge, many simulations opt for a practical approach by simplifying neutrino transport. They achieve this by utilizing a limited set of the neutrino angular distribution moments~\cite{Shibata:2011kx, Cardall:2012at, thorne1981relativistic}. These moments effectively encapsulate the essential characteristics of the neutrino angular distribution, enabling a more computationally manageable treatment while retaining crucial information.

It has been demonstrated that  one can still  harness these limited information (existing in the neutrino moments) for assessing the occurrence  of $\nu$ELN crossings in CCSN and NSM simulations. While the initial efforts were focused on analytical/semi-analytical techniques~\cite{Dasgupta:2018ulw, Abbar:2020fcl, Johns:2021taz, Johns:2019izj, Nagakura:2021hyb, Richers:2022dqa, Nagakura:2021suv}, which are limited by either their computational speeds or their inefficiencies, 
it has  been recently shown that machine learning (ML) techniques can do a great job in capturing  FFCs in CCSN and NSM  simulations~\cite{Abbar:2023kta, Abbar:2023zkm}. 
 While ML methods demand substantial data and an initial training phase, their post-training speed and efficiency are remarkable. This provides a promising avenue for real-time FFC detection within CCSN and NSM simulations.
This becomes particularly critical considering the scale of FFCs. Integrating them into simulations necessitates the development of sub-grid models for quantum kinetic equations. These models involve performing neutrino-radiation hydrodynamic simulations using classical neutrino transport, while treating flavor conversions as a crucial sub-grid aspect. Capturing FFCs would be a crucial part of such sub-grid models~\cite{Nagakura:2023jfi, Abbar:2023ltx}.
Furthermore, the integration of pre-trained ML models follows a straightforward procedure, substantially reducing the need for extensive coding work, even during post-processing analysis of FFC occurrences. ML techniques not only offer the quickest and most precise means to detect FFCs but also hold potential for improved performance when trained in progressively complex environments.

While all the aforementioned methods  have been successfully employed to capture FFCs, they all faces a critical limitation: they exclusively detect axisymmetric $\nu$ELN crossings. Simply put, they are not sensitive to the non-radial moments, which are sensitive to asymmetries in $\phi_\nu$~\footnote{ Note that the approach outlined in Ref.~\cite{Abbar:2020fcl} has, in principle, the capability to extract information present in the non-radial moments. However, its analytical nature causes it to operate at a slower pace in practice.}(see Sec.~\ref{sec:models} for the detailed definition of the neutrino moments). 
This limitation is partly rooted in the significant lack of comprehensive information regarding the neutrino angular distribution shape in $\phi_\nu$, with almost no analytical understanding of its  form.
This limitation carries substantial weight, especially when considering the prevalence of non-axisymmetric $\nu$ELN  crossings within the SN  environment~\cite{Nagakura:2019evv}. Hence, any methodology solely sensitive to axisymmetric $\nu$ELN crossings risks overlooking a considerable portion of the crossings.
Moreover, this issue is intricately linked to the significance of non-axisymmetric features within rotating SN models, 
given the possible 
 correlation between non-axisymmetric crossings and non-radial fluxes~\cite{Nagakura:2019sig}.
 Regarding this issue, it is illuminating to  note that the non-axisymmetric crossings never happen without non-radial fluxes.

In this study, we aim to overcome this constraint by developing ML techniques capable of identifying non-axisymmetric $\nu$ELN crossings. Specifically, our ML methods exhibit sensitivity to all neutrino moments available in both radial and non-radial directions. To accomplish this, we utilize non-rotating axisymmetric and rotating CCSN  models~\cite{Nagakura:2019evv, Harada:2021ata} to train and validate our ML models.
Our findings demonstrate that our ML models not only achieve significant accuracy in detecting both non-axisymmetric and axisymmetric crossings but also effectively differentiate between these two categories of $\nu$ELN crossings.

In the upcoming section, we discuss our CCSN models, the source of our data. In Sec.~\ref{sec:ML}, we introduce our ML algorithms and we  assess their performance  in detecting  $\nu$ELN crossings in both rotating and non-rotating models. Finally and before presenting our conclusions, we develop ML models capable of differentiating between axisymmetric and non-axisymmetric $\nu$ELN crossings.

  \section{CCSN models}\label{sec:models}

The training data in this paper is taken from Boltzmann-neutrino-radiation hydrodynamics simulations. The code simultaneously solves the Boltzmann equation for neutrino transport \cite{Sumiyoshi:2012za}, the Poisson equation for self-gravity, and the hydrodynamics equations. In order to realize the neutrino trapping in infalling matter by the beaming effect, the special relativistic effects are implemented by using a two-grid approach \cite{Nagakura:2014nta}. A moving-mesh approach is employed to track the proper motion of the proto-neutron star (PNS) \cite{Nagakura:2017moving}. For the simulations, the Furusawa-Togashi equation of state (FT EOS) based on the multi-nuclear variational method is adopted \cite{Furusawa:2017auz}. The neutrino-matter interactions are carefully computed to be consistent with the FT EOS.
  

We computed several angular moments and the classes of crossing from the neutrino radiation field. The zero-th moment $I_0^\nu\;(\nu \in (\nu_{\rm e},\bar{\nu}_{\rm e}))$, the number density, is defined by
\begin{equation}
I_0^{\nu} = \int d\Omega \int_0^\infty \frac{E_\nu^2 dE_\nu}{(2\pi)^3} f_\nu(\boldsymbol p).
\end{equation}
Similarly, the first moment $I_\beta^\nu\;(\nu \in (\nu_{\rm e},\bar{\nu}_{\rm e});\,\beta\in(r,\theta,\phi))$, the number flux, is defined by
\begin{equation}
I_\beta^\nu = \int d\Omega \int_0^\infty \frac{E_\nu^2 dE_\nu}{(2\pi)^3} n_\beta f_\nu(\boldsymbol p),
\end{equation}
where $n_r=\cos\theta_\nu$, $n_\theta=\sin\theta_\nu \cos\phi_\nu$, and $n_\phi=\sin\theta_\nu \sin\phi_\nu$. Note that these quantities are evaluated in the fluid rest frame. Furthermore, we define the axisymmetrized ELN by
\begin{equation}
\bar{G}(\mu) = \int_0^{2\pi}d\phi_\nu \int_0^\infty \frac{E_\nu^2 dE_\nu}{(2\pi)^3} (f_{\nu_{\rm e}}(\boldsymbol p) - f_{\bar{\nu}_{\rm e}}(\boldsymbol p)).
\end{equation}
With the axisymmetrized ELN, we classify the crossings: no crossing means $G(\boldsymbol v)$ has the same sign for all $\boldsymbol v$, crossing means $\bar{G}(\mu)$ has both positive and negative parts, and non-axisymmetric crossing means $\bar{G}(\mu)$ has the same sign for all $\mu$ but $G(\boldsymbol v)$ has both positive and negative parts.

\begin{figure} [tb]
\centering
\begin{center}
\includegraphics*[width=0.45\textwidth]{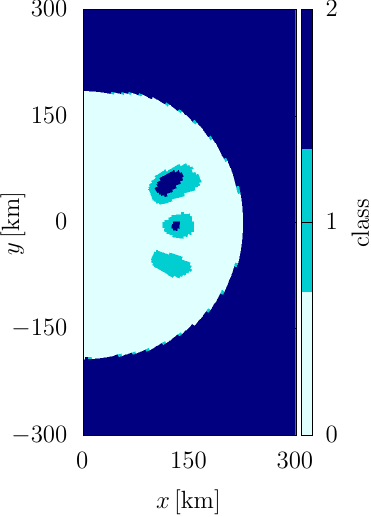}
\end{center}
\caption{
The distribution of the crossing types for the rotating model. Class 1, 2, and 3 indicate ``no crossing'', ``non-axisymmetric'', and ``crossing'', respectively. The time of this snapshot is $160\,{\rm ms}$.
}
\label{fig:ft15m4rad_crossings}
\end{figure}

We employed two models of supernova simulations, non-rotating and rotating ones. The non-rotating model is the same as that employed in Ref. \cite{Abbar:2023zkm}: the model with a progenitor mass of $11.2\,M_\odot$. For the training, we choose the snapshots at $200$, $250$, and $300\,{\rm ms}$ after the bounce. The rotating model is taken from Ref. \cite{Harada:2021ata}: the model with the progenitor mass of $15\,M_\odot$ and the initial $4\,{\rm rad\,s^{-1}}$ rotation. This rotating model presents the rotation-induced FFC: the equatorially extended matter profile by centrifugal force enhances the $\nu_{\rm e}$ absorption to create the ELN crossing. Furthermore, some regions of rotation-induced FFC show non-axisymmetric crossing, as shown in figure \ref{fig:ft15m4rad_crossings}. The outer region with the ``crossing'' class is caused by the inward coherent scattering, and the inner ``crossing'' and ``non-axisymmetric'' regions indicate the rotation-induced FFC regions. We choose the snapshots at every $10\,{\rm ms}$ from $100\,{\rm ms}$ to $200\,{\rm ms}$ after the bounce.

   \section{ML algorithms}\label{sec:ML}

 ML is a branch of artificial intelligence focused on empowering computers to learn and improve from experience without being explicitly programmed, enabling them to recognize patterns, make predictions, and derive insights from data.
As discussed before, recent progress has shown the successful application of ML algorithms in identifying $\nu$ELN crossings within simulations of CCSNe and NSMs~\cite{Abbar:2023kta, Abbar:2023zkm}. To initiate, we offer a concise introduction to relevant ML methodologies for this work. Following this, we provide an extensive exploration of our research findings.

In order to  training and testing of our ML models, we utilize neutrino moments derived from the CCSN simulations detailed in the preceding section. These moments serve as input data to train the ML models in distinguishing the presence or absence of the $\nu$ELN crossings.
In addition and in order to improve the efficiency of our ML algorithms, we do a feature engineering which is similar in spirit
to what performed in Ref.~\cite{Abbar:2023kta, Abbar:2023zkm}. Indeed instead of considering the original 
neutrino angular moments, we use 
\begin{equation}
\alpha = I^{\bar\nu_e}_0/I^{\nu_e}_0,\  F_{\beta}^{\nu_e} = I^{\nu_e}_\beta/I^{\nu_e}_0,\ \mathrm{and}\ F_{\beta}^{\bar\nu_e} = I^{\bar\nu_e}_\beta/I^{\bar\nu_e}_0,
\end{equation}
as the relevant features to be considered in the ML algorithms. Here $\beta \in (r,\theta, \phi)$,  resulting in a total of 7 input features for our ML models.
Such a  feature engineering is justified by bearing in mind that an overall normalisation factor does not affect the occurrence
of  $\nu$ELN crossings.
It's also worth highlighting that our current emphasis is primarily on the first two moments. These moments are of particular interest as they are the ones typically tracked directly in the simulation processes.

It's also worth noting the potential for data augmentation in this context, particularly due to the interchangeability of $\nu$ and $\bar\nu$, as well as the rotational flexibility of $I_\beta$ within different reference frames, while maintaining the occurrence of ELN crossings. To preserve the original distribution obtained from SN simulations, we selectively generate artificial data as a fraction of the initial dataset. Also constraints are applied to limit the rotation angle around the $\theta$ and $\phi$ axes within the range of $(0, \pi/2)$, preventing significant modification that could dominate the distribution in the backward direction.

In order to unbiasedly evaluating our ML algorithms, it is essential to partition the dataset into three distinct sets, namely 
the training , the validation , and the  test sets. These sets are used for training the ML algorithms, for fine-tuning the algorithm's hyperparameters, and for assessing the ML method's performance on previously unseen data, respectively.

To thoroughly evaluate the performance of our ML algorithm, we consider the precision and recall metrics, defined as follows:
\begin{equation}
\begin{split}
&\mathrm{accuracy} = \frac{T_p + T_n}{T_p + T_n + F_p + F_n} \\
&\mathrm{precision} = \frac{T_p}{T_p+F_p} \\
&\mathrm{recall} = \frac{T_p}{T_p+F_n} \\ 
&F_1 = 2\times \frac{\mathrm{precision} \times \mathrm{recall} }{\mathrm{precision} + \mathrm{recall}},
\end{split}
\end{equation}
with $T(F)_{p(n)}$ denoting True (False) positive (negative) classifications.
A discerning reader will notice that the precision/recall metric informs us about the reliability/detectability of  classifications, while $F_1$ is their harmonic mean. In this study, we opt for accuracy as the suitable metric because we aim for equal sensitivity to the presence or absence of $\nu$ELN crossings.

This study examines key ML algorithms:
i) Logistic Regression (LR): a statistical classifier ideal for binary outcomes,
ii) k-Nearest Neighbors (KNN): an intuitive algorithm that classifies based on nearest neighbors,
iii) Support Vector Machine (SVM): powerful for class separation using hyperplanes in high-dimensional spaces, and
iv) Decision Tree (DT): for classification and regression by feature-based data splitting.
Two nuances merit clarity for LR and SVM:
LR, despite its linear operation, requires preprocessing for nonlinear $\nu$ELN crossings detection.
SVM employs the radial basis function (RBF) kernel with $\gamma = 100$. For more discussions on these issues and the 
different ML algorithms consult with Refs.~\cite{Abbar:2023kta, Abbar:2023zkm}.

Our forthcoming section will showcase the results regarding the performance of our ML models. To foster transparency and collaboration, our methodologies are accessible on \href{?}{GitHub}.

\subsection{ML performance on the rotating SN model}

Prior to describing the intricacies of non-axisymmetric crossings, our foremost aim is to assess the efficacy of the pre-trained ML models as detailed in Ref.~\cite{Abbar:2023zkm}. These models were initially trained using data derived from the non-rotating supernova (SN) model. Now, we seek to ascertain their performance when applied to the rotating core-collapse supernova (CCSN) model.
It's crucial to underline that our analysis in this subsection is specifically focused on axisymmetric crossings that have been well-documented in existing literature (so we only use  $\alpha$, $F_r^{\nu_e}$, and $F_r^{\bar\nu_e}$ as the inputs of ML models). Given the differing physics of rotating and non-rotating models, this assessment provides valuable insights into how ML performs on entirely unanticipated data.

The summary of the metric scores for our pre-trained ML  algorithms (described in Ref.~\cite{Abbar:2023zkm}), tested on the rotating  SN model can be found in Table.~\ref{tab:oldML}. Notably, the pre-trained ML models exhibit outstanding performance when applied to the rotating SN models, where the  accuracies $> 90\%$  can be easily reached. This strongly supports the notion that ML techniques can be confidently employed for the evaluation of FFC occurrences in CCSNe.

 \begin{table}[tb]
\centering
\caption{A summary of the metric scores of the previously-trained ML algorithms in Ref.~\cite{Abbar:2023zkm}, when tested on the rotating SN model. Alongside each algorithm, one can find its corresponding accuracy score.}
\begin{tabular}[t]{|lcc|c|}
\hline
& \textcolor{black}{ \textbf{{LR} (n = 2)} (95\%)}    \\
\hline
& precision & recall & $F_1$-score \\
\hline
no  crossing & 98\% & 95\% & 96\% \\
 crossing&90\%&97\% & 94\% \\
\hline
&\textcolor{black}{  \textbf{{KNN (n=3)}} (93\%)  } \\
\hline
&precision&recall & $F_1$-score \\
\hline
no  crossing&95\%&95\%&95\%\\
 crossing&90\%&90\%&90\%\\
\hline
&\textcolor{black}{   \textbf{{SVM}} (92\%)}\\
\hline
&precision&recall & $F_1$-score\\
\hline
no  crossing&92\%&96\%&94\%\\
 crossing&92\%&85\%&88\%\\
\hline
&\textcolor{black}{   \textbf{{DT}} (94\%) }\\
\hline
&precision&recall & $F_1$-score \\
\hline
no  crossing&94\%&97\%&95\%\\
 crossing&94\%&87\%&91\%\\
\hline
\end{tabular}
\label{tab:oldML}
\end{table}%

\subsection{ML performance in the presence of non-axisymmetric crossings}\label{sec:nonax}

\begin{figure} [tb]
\centering
\begin{center}
\includegraphics*[width=.45\textwidth, trim= 0 0 30 20, clip]{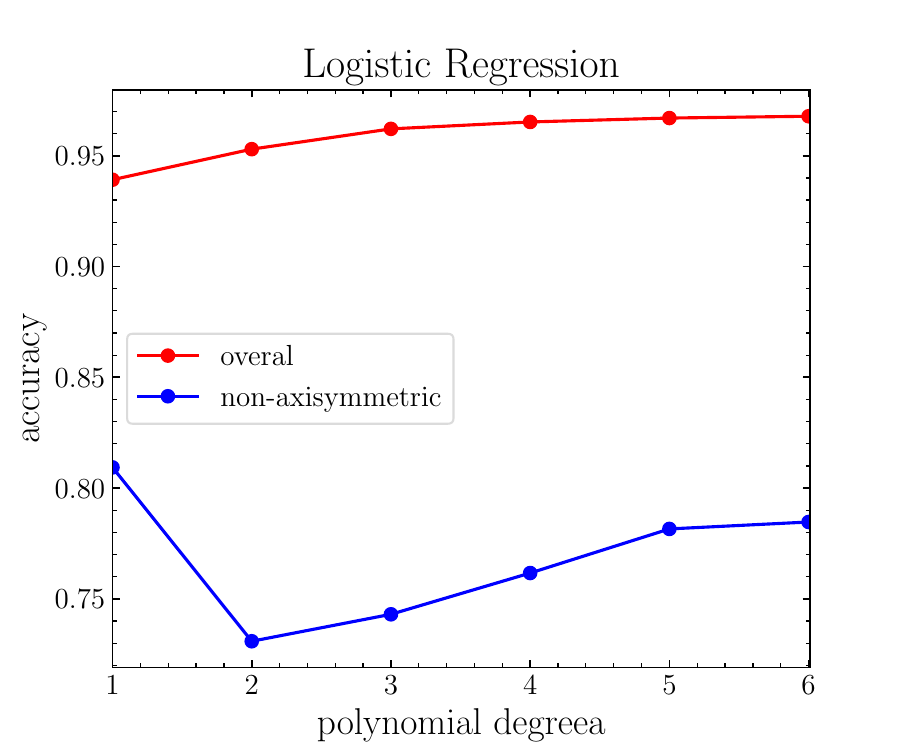}
\end{center}
\caption{
The accuracy of the LR algorithm when evaluated on a dataset comprising both rotating and non-rotating models, as a function of the polynomial degree  of nonlinear transformations. The red curve represents the overall accuracy across the entire dataset, while the blue curve represents the recall score specifically for non-axisymmetric crossings.}
\label{fig:LR}
\end{figure}

\begin{table}[b]
\centering
\caption{A summary of the metric scores of the  ML algorithms trained on both rotating and non-rotating SN models. Alongside each algorithm, one can find its corresponding accuracy score.
Note that the scores for the non-axisymmetric crossings are calculated by inferring whether each point known to be 
 a non-axisymmetric crossing (from the actual angular distributions) is classified as crossing or not. 
 This process differs from the three-lass classification discussed in Sec.~\ref{sec:3c}.
}

\begin{tabular}[t]{|lcc|c|}
\hline
& \textcolor{black}{ \textbf{{LR} (n = 2)} (95\%)}    \\
\hline
& precision & recall & $F_1$-score \\
\hline
no  crossing & 97\% & 95\% & 96\% \\
 crossing&93\%&96\% & 94\% \\
 non-axisymmetric&100\%&73\% & 84\% \\
\hline
&\textcolor{black}{  \textbf{{KNN (n=3)}} (100\%)  } \\
\hline
&precision&recall & $F_1$-score \\
\hline
no  crossing&100\%&100\%&100\%\\
 crossing&100\%&100\%&100\%\\
 non-axisymmetric&100\%&99\% & 100\% \\
\hline
&\textcolor{black}{   \textbf{{SVM}} (95\%)}\\
\hline
&precision&recall & $F_1$-score\\
\hline
no  crossing & 97\% & 95\% & 96\% \\
 crossing&93\%&96\% & 94\% \\
 non-axisymmetric&100\%&73\% & 84\% \\
\hline
&\textcolor{black}{   \textbf{{DT}} (100\%) }\\
\hline
&precision&recall & $F_1$-score \\
\hline
no  crossing&100\%&100\%&100\%\\
 crossing&100\%&99\%&100\%\\
 non-axisymmetric&100\%&99\% & 100\% \\
\hline
\end{tabular}
\label{tab:newML}
\end{table}%

In this part, we evaluate the performance of our ML models after training them on datasets containing both rotating and non-rotating  SN models. It's important to note that while we include non-axisymmetric crossings in our analysis, we do not attempt to distinguish them from the axisymmetric ones.

In Fig.~\ref{fig:LR}, we present the accuracy of the LR algorithm as a function of the polynomial degree of nonlinear transformations.
As illustrated, regardless of the polynomial degree used, more than 75\% of the  non-axisymmetric crossings are accurately captured, with a remarkable  overall accuracy.

In Table~\ref{tab:newML}, we present the metric scores for different ML algorithms, trained and tested  on datasets containing both rotating and non-rotating  SN models. 
It is readily apparent that these scores are exceptional, reaching even a 100\% accuracy for  the KNN and DT algorithms. This remarkable performance aligns with findings in Refs.~\cite{Abbar:2023kta, Abbar:2023zkm}, which highlighted that the primary source of error in capturing $\nu$ELN crossings (in those references) arises from the variations in the assumed \emph{forms} of the neutrino angular distributions. To clarify, when dealing with various forms of the neutrino angular distributions, we introduce noisy labels into the supervised learning process. This introduces the possibility that, at the same point in the parameter space, we may encounter instances where both the presence and absence of crossings are observed.

Notably,  in this study we have eliminated the possibility for such an apparent variations in the form of the neutrino angular distributions since all the information comes directly from simulation data. For this reason, such exceptional accuracies reported in Table~\ref{tab:newML} are accissible.
However, it's important to anticipate a potential decrease in accuracy when our ML methods are applied to unseen data. Such an expected decrease may arise from variations in the approximations employed in the neutrino transport, resulting in differing effective shapes of neutrino angular distributions.

\subsection{Performance of ML in the neutrino decoupling region}

\begin{table}[tbh]
\centering
\caption{A summary of the metric scores of our  ML algorithms within the neutrino decoupling region, which we consider to be $r\lesssim 100$~km. Alongside each algorithm, one can find its corresponding accuracy score.}
\begin{tabular}[t]{|lcc|c|}
\hline
& \textcolor{black}{ \textbf{{LR} (n = 2)} (95\%)}    \\
\hline
& precision & recall & $F_1$-score \\
\hline
no  crossing & 97\% & 95\% & 96\% \\
 crossing&93\%&96\% & 94\% \\
\hline
&\textcolor{black}{  \textbf{{KNN (n=3)}} (100\%)  } \\
\hline
&precision&recall & $F_1$-score \\
\hline
no  crossing&100\%&100\%&100\%\\
 crossing&100\%&100\%&100\%\\
\hline
&\textcolor{black}{   \textbf{{SVM}} (95\%)}\\
\hline
&precision&recall & $F_1$-score\\
\hline
no  crossing & 97\% & 95\% & 96\% \\
 crossing&93\%&96\% & 94\% \\
\hline
&\textcolor{black}{   \textbf{{DT}} (100\%) }\\
\hline
&precision&recall & $F_1$-score \\
\hline
no  crossing&100\%&100\%&100\%\\
 crossing&100\%&100\%&100\%\\
\hline
\end{tabular}
\label{tab:tab_dec}
\end{table}%

In the preceding part, we conducted a comprehensive evaluation of   ML performance within the  SN environment. It is crucial to emphasize that FFCs are anticipated to exert the most substantial influence when they transpire in the proximity to the surface of the PNS, specifically within the SN post-shock zone. This can be attributed to two fundamental factors. Firstly, FFCs occurring in deeper SN regions can potentially result in broader and more profound $\nu$ELN crossings, consequently leading to more pronounced flavor conversions~\cite{martin2020dynamic, bhattacharyya2020late, bhattacharyya2021fast, wu2021collective,richers2021neutrino,zaizen2021nonlinear,richers2021particle,bhattacharyya2022elaborating,grohs2022neutrino,abbar2022suppression,richers2022code,zaizen2023simple}. Moreover, any flavor conversion that takes place above the SN shock is not anticipated to significantly impact the  CCSN dynamics~\cite{Ehring:2023abs, Ehring:2023lcd}, although it may still affect the neutrino signal.

Considering this, in addition to evaluating our ML methods' overall performance, we specifically examined their performance in SN regions located well below the shock. To do this, we assessed the performance of our ML algorithms in SN zones where the radial distance  is $\lesssim 100$~km. As illustrated in Table~\ref{tab:tab_dec}, our ML models still effectively capture $\nu$ELN crossings within such deep regions.

\subsection{Axisymmetric vs non-axisymmetric $\nu$ELN crossings }\label{sec:3c}

\begin{table}[b]
\centering
\caption{A summary of the metric scores of the  ML algorithms trained on both rotating and non-rotating SN models. Alongside each algorithm, one can find its corresponding accuracy score.}
\begin{tabular}[t]{|lcc|c|}
\hline
&\textcolor{black}{  \textbf{{KNN (n=3)}} (99\%)  } \\
\hline
&precision&recall & $F_1$-score \\
\hline
no  crossing&100\%&100\%&100\%\\
 crossing&100\%&100\%&100\%\\
 non-axisymmetric&93\%&91\% & 92\% \\
\hline
&\textcolor{black}{   \textbf{{DT}} (100\%) }\\
\hline
&precision&recall & $F_1$-score \\
\hline
no  crossing&100\%&100\%&100\%\\
 crossing&100\%&100\%&100\%\\
 non-axisymmetric&93\%&92\% & 92\% \\
\hline
\end{tabular}
\label{tab:multiclass}
\end{table}%

In the previous sections, we did not make an effort to differentiate between axisymmetric and non-axisymmetric $\nu$ELN crossings. However, doing so can hold significant implications for various reasons.
On the one hand, this information is crucial to develop our understanding of the different types of the $\nu$ELN crossings. For instance by differentiating between these two types of crossings one can
perform a systematic investigation of the correlation between non-axisymmetric crossings and non-radial fluxes in the SN simulations. In addition, such information is useful and important for predicting the asymptotic outcome of FFCs. 
As a result, we now endeavor to determine whether it is possible, in principle, to distinguish between axisymmetric and non-axisymmetric $\nu$ELN crossings.

To achieve this distinction, a three-class classification algorithm must be developed. This is distinct from previous scenarios where the task was to simply determine the presence or absence of crossings. 
The outcomes of this three-class classification are outlined in Table~\ref{tab:multiclass}. Remarkably, both the KNN and DT algorithms excel in this classification task. Conversely, the SVM and LR fail to produce satisfactory results and are not shown in Table~\ref{tab:multiclass}.

It's important to recognize that the results presented in Table~\ref{tab:multiclass} are fundamentally distinct from those discussed in Table~\ref{tab:newML}. Therein, we exclusively assessed the performance of our two-class classification ML algorithm in identifying non-axisymmetric crossings, assuming we already have information about the crossing types. However, the results presented here represent a genuine attempt by the ML algorithm to differentiate between different three classes, a capability that can prove useful when information regarding the nature of the crossings is not available.

\section{DISCUSSION AND OUTLOOK}

Recent reseraches have highlighted the  capabilities of ML in identifying axisymmetric $\nu$ELN crossings in simulations of CCSNe and NSMs~\cite{Abbar:2023kta, Abbar:2023zkm}.
This study represents a pioneering effort in capturing non-axisymmetric $\nu$ELN crossings in CCSNe by employing conventional ML algorithms. These algorithms utilize all accessible neutrino angular moments, particularly those that exhibit sensitivity to asymmetries in $\phi_\nu$.
To train and test our model, we utilized neutrino angular distributions from both rotating and non-rotating CCSN models. Our results demonstrate the remarkable efficacy of ML models in detecting both axisymmetric and non-axisymmetric $\nu$ELN crossings.

An essential characteristic of ML algorithms is their ability to generalize effectively to new, unseen data. To assess the generalizability of our ML models, we evaluate the performance of an ML model trained on data collected from the non-rotating SN model when applied to data from the rotating  one. This evaluation is particularly crucial as these two datasets could be considered somewhat unrelated. Our findings demonstrate that the ML models excel in terms of generalization, showcasing their ability to perform well even in scenarios where they have not been previously exposed.

In our analysis encompassing rotating and non-rotating models, we have developed ML models capable of achieving remarkably high accuracies in detecting both axisymmetric and non-axisymmetric $\nu$ELN crossings. Our models achieve an overall accuracy exceeding $90\%$. We have also evaluated the performance of these ML models in the post-shock region of supernovae (SN).
Our findings reveal the notable capability of ML methods, particularly in the deepest SN regions, specifically when $r\lesssim100$~km.

In addition to capturing both axisymmetric and non-axisymmetric $\nu$ELN crossings, we have also demonstrated the capability of ML models to discern between these  crossing types. We specifically highlight the achievement of differentiation accuracy exceeding 90\%. This distinct ability of our ML models holds significant importance, given 
the usefulness of such a distinction in predicting the asymptotic outcome of FFCs, and also in 
 investigation of the correlation between non-axisymmetric crossings and non-radial fluxes in the SN simulations. 

 In essence, our study marks a pioneering endeavour in capturing non-axisymmetric $\nu$ELN crossings within CCSN simulations. However, a significant limitation of our work stems from focusing solely on data derived from CCSN models. Considering the pronounced importance of FFCs  in NSMs and the  disparity in the geometry between these distinct astrophysical scenarios, a crucial enhancement to our study involves the training and development of ML models using NSM data.
By doing so, we can achieve a more comprehensive and robust detection of $\nu$ELN crossings in both CCSNe and NSMs. Needless to say, taking into account such measures improves the performance of ML models in detecting FFCs in simulations of both CCSNe and NSMs.


\section*{Acknowledgments}
We would like to thank Georg Raffelt for useful discussions. 
We would also like to express our sincere gratitude to the Institute of Physics of Academia Sinica for their warm hospitality and support during the \emph{Focus Workshop on Collective Oscillations and Chiral Transport of Neutrinos}, where the inception of this project took place. Their gracious hosting and collaborative environment played a pivotal role in shaping the foundation of our work.
S.A. was supported by the German Research Foundation (DFG) through
the Collaborative Research Centre  ``Neutrinos and Dark Matter in Astro-
and Particle Physics (NDM),'' Grant SFB-1258, and under Germany’s
Excellence Strategy through the Cluster of Excellence ORIGINS
EXC-2094-390783311.
A. H. was supported by JSPS KAKENHI Grant Numbers JP20H01905 and JP21K13913.
H. N. was supported by Grant-in-Aid for Scientific Research (23K03468) and also by the NINS International Research Exchange Support Program. The numerical simulations for CCSNe were carried out by using "K", ”Fugaku”, and the highperformance computing resources of ”Flow” at Nagoya University ICTS through the HPCI System Research Project (Project ID: 220173, 220047, 220223, 230033, 230204, 230270).
We would also like to acknowledge the use of the following softwares: \textsc{Scikit-learn}~\cite{pedregosa2011scikit}, \textsc{Matplotlib}~\cite{Matplotlib}, \textsc{Numpy}~\cite{Numpy}, \textsc{SciPy}~\cite{SciPy}, and \textsc{IPython}~\cite{IPython}.

\bibliographystyle{elsarticle-num}
\bibliography{Biblio}

\clearpage

\end{document}